\documentclass[a4paper]{SIGEport}
\PassOptionsToPackage{table}{xcolor}
\usepackage{graphicx}
\usepackage{geometry}
\usepackage{cite}
\usepackage{amsmath}
\usepackage{url}
\usepackage{booktabs}
\usepackage{array}
\usepackage{tabularx}
\usepackage{arena-style}
\usepackage[hidelinks]{hyperref}

\newcommand{\HikariRecordCount}{16{,}009}

\newcommand{\HikariTimestampRate}{93.27\%}
\newcommand{\HikariOrderRate}{89.92\%}

\newcommand{\HikariSingletonEntityRate}{56.86\%}
\newcommand{\HikariSingletonSignatureRate}{52.68\%}
\newcommand{\HikariEmailPatternCount}{3}
\newcommand{\HikariUserPathPatternCount}{191}
\newcommand{\HikariGlobalIpPatternCount}{10{,}177}
\newcommand{\SocpilotIncidentCount}{200}

\newcommand{\SocpilotEventCount}{2{,}249}
\newcommand{\SocpilotMappedTaskCount}{1{,}147}

\newcommand{\SocpilotPrivacyIssueCount}{0}
\newcommand{\SocpilotPolicyCatalogCoverage}{40\%}

\makeatletter
\def\resumo{\normalfont%
  \if@twocolumn%
    \@IEEEabskeysecsize\bfseries\textit{Abstract}---\,%
  \else%
    \begin{center}\vspace{-1.78ex}\@IEEEabskeysecsize\textbf{Abstract}\end{center}\quotation\@IEEEabskeysecsize%
  \fi\@IEEEgobbleleadPARNLSP}
\def\chave{\normalfont%
  \if@twocolumn%
    \@IEEEabskeysecsize\bfseries\textit{Keywords}---\,\relax%
  \else%
    \begin{center}\vspace{-1.78ex}\@IEEEabskeysecsize\bfseries Keywords\end{center}\quotation\@IEEEabskeysecsize%
  \fi\@IEEEgobbleleadPARNLSP}
\def\fnum@table{\scriptsize{TABLE~\thetable}}

\def\thebibliography#1{\section*{References}%
  \addcontentsline{toc}{section}{References}%
  \footnotesize \vskip 0.3\baselineskip plus 0.1\baselineskip minus 0.1\baselineskip%
  \list{\@biblabel{\@arabic\c@enumiv}}%
  {\settowidth\labelwidth{\@biblabel{#1}}%
  \leftmargin\labelwidth
  \advance\leftmargin\labelsep\relax
  \itemsep 0pt plus .5pt\relax%
  \usecounter{enumiv}%
  \let\p@enumiv\@empty
  \renewcommand\theenumiv{\@arabic\c@enumiv}}%
  \let\@IEEElatexbibitem\bibitem%
  \def\bibitem{\@IEEEbibitemprefix\@IEEElatexbibitem}%
  \def\newblock{\hskip .11em plus .33em minus .07em}%
  \if@technote\sloppy\clubpenalty4000\widowpenalty4000\interlinepenalty100%
  \else\sloppy\clubpenalty4000\widowpenalty4000\interlinepenalty500\fi%
  \sfcode`\.=1000\relax}

\makeatother

\newif\ifanonymous
\anonymoustrue
\ifdefined\anonymous
  \ifnum\anonymous=0
    \anonymousfalse
  \fi
\fi

\begin{document}

\title{From Production SIEM to\\Reusable Cybersecurity Artifacts}

\author{Sidnei Barbieri$ ^1 $, Leonardo Vaz de Meneses$ ^1 $, \'{A}gney Lopes Roth Ferraz$ ^1 $, \\ Wagner Comin Sonaglio$ ^1 $, and Louren\c{c}o Alves Pereira J\'{u}nior$ ^1 $\\
    {\small $ ^1 $Computer Science Division, Aeronautics Institute of Technology (ITA), S\~{a}o Jos\'{e} dos Campos/SP - Brazil}\\
    {\small \{sidneisb, leonardomeneses, roth, sonaglio, ljr\}@ita.br}}

\maketitle

\begin{resumo}
Operational evidence is not automatically scientific evidence. The most realistic Security Operations Center (SOC) data is production telemetry, yet it remains scientifically inaccessible because raw logs cannot be released; as a result, research relies on synthetic or dated datasets. We treat the boundary between private production telemetry and reusable research artifacts as the design object: a methodology that extracts, anonymizes, structures, and validates Security Information and Event Management (SIEM) data from a production financial SOC while preserving task-relevant investigative structure within a declared privacy boundary. Two consumers stress the same artifact. As training material, it fails loudly: 37 MITRE ATT\&CK-mapped HIKARI challenges work only when anonymization preserves temporal order and entity consistency. As a measurement substrate, it fails quietly: across 200 SOCpilot incidents, a deterministic verifier detects non-compliant Large Language Model (LLM) actions that are absent from the human baseline. The result is a measurable privacy--utility boundary rather than a formal anonymity claim.
\end{resumo}

\begin{chave}
SIEM telemetry, anonymization, privacy--utility boundary, reusable datasets, SOC
\end{chave}

\section{Introduction}

Detecting, investigating, and responding to modern threats depend on data that reflects what analysts actually see. This requirement is becoming sharper as cyber-ranges and Large Language Model (LLM)-assisted response systems move from demonstrations to evaluation substrates: both need realistic event sequences, benign background activity, and cross-source entities that survive correlation. Yet the datasets available for research and training are mostly synthetic, dated, or single-source, and the realistic alternative, production telemetry, cannot be shared without resolving privacy, traceability, and institutional-security constraints. The result is a persistent gap: detection methods, analyst training, and autonomous defense benchmarks often rely on data that does not reflect operational reality.

Production telemetry is operationally useful but scientifically inaccessible: it is the most realistic security data an organization holds, yet without a declared boundary between operational data and research artifacts, it never becomes reproducible science. This paper addresses that gap by treating the artifact boundary as a first-class design object and presents a methodology that converts production Security Information and Event Management (SIEM) telemetry into artifacts that preserve task-relevant investigative structure under a declared privacy boundary. Starting from a real financial-sector environment, we extract, anonymize, structure, and quality-control logs from heterogeneous sources: endpoint detection and response (EDR), firewall, web application firewall (WAF), intrusion prevention system (IPS), unified threat management (UTM), Active Directory (AD), proxy, and operating-system events. The pipeline preserves the benign-to-malicious balance and the investigative semantics that make a dataset useful. The contribution is an auditable workflow that converts production SIEM logs into anonymized JavaScript Object Notation (JSON) datasets with data dictionaries and quality controls; a playbook-oriented dataset design mapped to MITRE ATT\&CK~\cite{mitre_attack}, preserving an 81.42\%/18.58\% benign-to-malicious distribution instead of an artificially attack-dense dataset; and a two-front validation showing that the treated datasets still support 37 HIKARI training challenges and a 200-incident SOCpilot evaluation with 800 LLM trajectories against human baselines.

\section{Background}
\label{sec:background}

Public intrusion datasets such as CICIDS-style captures~\cite{cicids2017}, UNSW-NB15~\cite{unswnb15}, CTU-13~\cite{ctu13}, LANL enterprise traces~\cite{lanl2015}, and CERT insider-threat data~\cite{cert2013} advanced reproducible evaluation, but each covers only part of the SOC setting: network flows, authentication graphs, insider narratives, or synthetic testbed activity. Production SOC investigation spans endpoint, perimeter, identity, and application logs and depends on event ordering, field semantics, and the presence of malicious activity within a realistic, benign background. Host-centric datasets closer to SOC incident narratives, such as DARPA Transparent Computing/OpTC~\cite{darpa_tc} and Splunk BOTS~\cite{splunk_bots}, capture richer endpoint provenance but remain single-program exercises rather than continuous, multi-source production telemetry with explicit privacy treatment. The gap is as much about \emph{provenance and structure} as \emph{volume}: the data a SOC trusts looks like production telemetry rather than a single-source capture.

A second line of work turns raw logs into structured, analyzable representations. Automated log parsing and benchmarking toolkits~\cite{logpai} show that a log's analytical value lies less in retaining every field than in exposing the attributes that support detection and correlation. We adopt this view at the dataset level: extraction is paired with deliberate feature selection and a per-source data dictionary, so that an anonymized record remains interpretable rather than an opaque token stream.

The realistic source, however, is production telemetry, which cannot be released without privacy-preserving treatment that keeps the data useful. Classical privacy models such as $k$-anonymity~\cite{kanonymity} and its $\ell$-diversity and $t$-closeness refinements protect identities by generalizing or suppressing quasi-identifiers, but applied naively to security logs, they erase exactly the fields an investigation needs: consistent host and account identities, command lines, and the communication structure of an attack. Our treatment is therefore semantic-preserving, not identity-erasing. It consistently pseudonymizes, preserves technical fields, and validates the result as an operational artifact for both cyber-range training and LLM-assisted incident-response evaluation. This complements the adversary emulation and policy verification lines within the same program~\cite{autosut,socpilot}.

\section{Methodology and Datasets}
\label{sec:method}

Figure~\ref{fig:pipeline} shows the boundary this methodology builds: private production telemetry is anonymized and structured, a utility and linkage-risk gate decides what may cross, and only the artifact that passes is consumed by the two validation fronts.

\begin{figure}[t]
\centering
\begin{tikzpicture}[arena, font=\footnotesize, node distance=4mm]
  \node[arena attack, text width=6.6cm] (src) {\textbf{Private production SIEM} \\\footnotesize endpoint · firewall · WAF · IPS · UTM · AD · proxy · OS\\\footnotesize raw logs cannot leave};
  \node[arena node, below=of src, text width=6.6cm] (treat) {Semantic-preserving anonymization · JSON structuring · per-source data dictionary};
  \node[arena policy, draw=arenaInk, line width=1pt, below=5mm of treat, text width=6.6cm] (gate) {\textbf{Artifact boundary}\\\footnotesize utility gate $\wedge$ linkage-risk gate\\\footnotesize only what passes crosses};
  \node[arena node, below=5mm of gate, text width=6.6cm] (art) {\textbf{Reusable research artifact}\\\footnotesize canonical packages · data dictionaries · manifest};
  \node[arena policy, below=8mm of art, text width=3.0cm, xshift=-1.75cm] (hikari) {HIKARI\\\footnotesize 37 challenges\\\emph{fails loudly}};
  \node[arena policy, below=8mm of art, text width=3.0cm, xshift=1.75cm] (soc) {SOCpilot\\\footnotesize 200 incidents\\\emph{fails quietly}};
  \draw[arena flow strong] (src) -- (treat);
  \draw[arena flow strong] (treat) -- (gate);
  \draw[arena flow strong] (gate) -- (art);
  \coordinate (bY) at ($(art.south)+(0,-4mm)$);
  \draw[arena flow strong, -] (art.south) -- (bY);
  \draw[arena flow strong, -] (hikari.north|-bY) -- (soc.north|-bY);
  \draw[arena flow strong] (hikari.north|-bY) -- (hikari.north);
  \draw[arena flow strong] (soc.north|-bY) -- (soc.north);
\end{tikzpicture}
\caption{Production-to-artifact boundary.}
\label{fig:pipeline}
\end{figure}

The treatment proceeds in seven stages: source selection, representative extraction, semantic-preserving anonymization, JSON structuring, feature selection, playbook organization, and utility validation. Treated this way, the datasets are engineered artifacts rather than raw exports.

Logs are extracted from the production SIEM across heterogeneous sources and processed using a semantically preserving strategy that controls sensitive information while preserving the relationships an investigation needs. The current HIKARI transform is a deterministic, rule-based replacement dictionary, not a salted hash or a cryptographic anonymizer: predefined terms and field patterns map to standardized semantic classes such as nominal account, privileged account, domain, application, and organization. The private dictionary provides within-release consistency and never crosses the release boundary. Technical fields are retained only when their investigative role outweighs linkage risk; consequently, globally routable addresses, rare commands, paths, and account-like strings are scanned rather than presumed safe. A public release would replace linkable identifiers with per-release-keyed tokens stored outside the artifact and apply a topology-preserving network transform when flow structure is required. Crypto-PAn~\cite{cryptopan} supplies the prefix-preserving baseline, but neither it nor formal cryptographic anonymity is claimed for the current candidate.

Extraction is purposive, not exhaustive. Events are added to the dataset when they meet five criteria: security relevance, source diversity, investigative potential, playbook alignment, and privacy feasibility. Each source is then converted to JSON, retaining only the attributes that carry detection value rather than all available fields. Following the log-analysis literature on structured representations~\cite{logpai}, features are kept when they provide semantic discrimination, appear consistently enough to support analysis, and matter to response. A per-source data dictionary assigns an interpretable meaning to every retained field, and benign-versus-malicious labels are derived from playbook-based rules on event name, technique, signature, action, indicator-of-compromise (IOC) reputation, frequency, and flow context, rather than ad hoc line-by-line tagging. The result preserves the complexity, noise, and ambiguity of real SOC investigation while keeping the controlled balance reproducible.

Table~\ref{tab:inventory} lists the eight anonymized datasets, totaling 16{,}009 records, along with their record and feature counts and investigative focus. Source diversity across endpoint, web/proxy, identity, and perimeter enables a single dataset to represent multiple investigative perspectives, more closely aligned with SOC practice than single-source data. Feature counts vary by source for a reason: endpoint (EDR) records keep 29 fields because process lineage, command line, hash, and ATT\&CK mapping all matter to an investigation, whereas perimeter records need far fewer.

\begin{table}[!htb]
\caption{Representative anonymized datasets}
\label{tab:inventory}
\centering\small
\arenazebra
\begin{tabularx}{\columnwidth}{@{}lrr>{\raggedright\arraybackslash}X@{}}
\toprule
\textbf{Dataset} & \textbf{Records} & \textbf{Feat.} & \textbf{Investigative focus} \\
\midrule
\path{utm}         & 7{,}999 & 14 & Perimeter filtering, denied traffic \\
\path{alsd}        & 4{,}564 & 21 & Web/proxy, navigation policy \\
\path{linuxOS}     &   969 & 13 & Operating-system commands, reverse shell, backdoor \\
\path{bloodhound}  &   747 & 13 & AD enumeration, lateral movement \\
\path{ips}         &   553 & 13 & Scanning, exploit attempts \\
\path{eap}         &   525 & 11 & Web-app attacks: SQL (Structured Query Language) injection, cross-site scripting \\
\path{edr}         &   389 & 29 & Endpoint, malware execution \\
\path{rbmws}       &   263 & 20 & Internal suspicious movement \\
\midrule
\textbf{Total}     & \textbf{16{,}009} & --- & 8 sources, 4 SOC investigation layers \\
\bottomrule
\end{tabularx}
\end{table}

The eight datasets span the four layers a SOC investigation traverses. The \emph{perimeter} layer (\path{utm}, \path{ips}) records denied traffic, scanning, and exploit attempts at the network edge. The \emph{endpoint} layer (\path{edr}, \path{linuxOS}) captures process lineage, malware execution, reverse shells, and backdoor creation, and carries the most fields. The \emph{identity} layer (\path{bloodhound}) exposes Active Directory enumeration and the lateral-movement paths that follow a foothold. The \emph{application} layer (\path{alsd}, \path{eap}) covers web and proxy navigation policies, as well as web application attacks such as SQL injection and cross-site scripting. A realistic investigation rarely lives in one layer. A single challenge may correlate a perimeter alert, an endpoint detection, and an identity enumeration into one narrative, which is possible only because anonymization preserved entity consistency across the four sources, so the same host or account is recognizable wherever it appears.

Across the HIKARI datasets, 81.42\% of records are benign, and 18.58\% are malicious or suspicious. This balance is intentional, and it matters for behavioral realism, investigation quality, and evaluation validity: benign events make the malicious ones less artificial, force the analyst to establish what is normal before flagging the anomalous, and enable measurement of sensitivity and false-positive behavior. Neither an attack-only dataset nor a benign-only dataset would support meaningful training or evaluation.

Because anonymization replaces real values with synthetic tokens, each dataset family is paired with a data dictionary that, for each source, documents the relevant fields, their formats, example values, and investigative meaning, so analysts do not misinterpret anonymized records and new challenges can be built reproducibly. Source-specific structure is preserved. Endpoint (EDR) records keep process name, parent process, path, user, hash, command line, severity, and ATT\&CK tactic and technique; perimeter (IPS, UTM) records keep source and destination IP, port, protocol, signature, action, and severity. Crossing the private-to-public boundary is gated by a dual-review protocol: the SOC operator performs initial sanitization and operational review, and the research side independently reprocesses the exported records with automated personally identifiable information (PII) detection, forbidden-term checks, rare-token review, and manual inspection.

Every candidate artifact passes five checks before release: privacy inspection, schema validation, source-specific review, challenge-usability validation, and semantic-preservation review. The guiding test is the preservation of investigative utility, not the absence of sensitive data alone, since naive redaction would trade one against the other.

The implementation keeps three artifact tiers separate. The first tier is private staging: HIKARI source exports, challenge documents, and 200 SOCpilot raw incident exports remain inside the controlled workspace and are not released artifacts. The second tier is the derived research artifact: anonymized JSON datasets, data dictionaries, challenge manifests, canonical incident packages, and verifier inputs. The third tier is the public or reviewer-facing package: a vetted subset rebuilt from manifests, plus scripts that validate schema, counts, token policy, and release eligibility. This separation matters because a SOC dataset can be safe only at the right tier. Releasing raw exports would violate the privacy boundary, while releasing only aggregate statistics would prevent a reviewer from reproducing training or LLM evaluation claims.

The two derived artifacts expose different scientific surfaces. The HIKARI artifact is a log dataset: eight JSON families, 16{,}009 records, per-source dictionaries, ATT\&CK labels, and challenge manifests that specify which evidence must appear at each injection step. Its release manifest records the source family, record count, field list, anonymization policy, and hash for each dataset, so a reviewer can rebuild Table~\ref{tab:inventory}, verify counts, and replay ingestion without seeing the original export. The SOCpilot artifact is an incident dataset: 200 canonical packages, each with anonymized incident metadata, category, severity, extracted tasks, mapped human actions, policy inputs, and verifier outputs. It is not a dump of case-management records. Fields that identify people, locations, administrators, or internal infrastructure stay in private staging; only the canonical package crosses the boundary. This distinction prevents a common failure mode in operational-data papers: calling an aggregate result reproducible while withholding the actual evaluation object.

The artifact boundary also determines what the paper can and cannot claim. We claim semantic-preserving pseudonymization under a declared protocol, not formal anonymity. We claim that downstream consumers can use the treated data, not that labels are error-free. Consequently, the release protocol treats privacy and label quality as gates. Privacy gates scan for forbidden strings, unresolved e-mail or account formats, rare tokens that could fingerprint a person or system, and cross-source joins that might recover a private identity. Label gates, sample incidents, and log records for independent analyst review, compare playbook-derived labels against the retained evidence, and record disagreements rather than silently repairing them. We report the current validated artifact construction and downstream use; a public release must pass those gates before it is safe to distribute.

For a reviewer, the intended artifact path is therefore explicit. First, inspect the manifest and confirm that private staging is excluded. Second, rebuild the anonymized tables and verify the eight source-family counts and field dictionaries. Third, run the schema and token-policy validators, which should fail closed when a forbidden token class appears. Fourth, replay the HIKARI ingestion subset and verify that challenge evidence appears only at the declared release step. Fifth, recompute the SOCpilot metrics from canonical incident packages, action traces, and verifier outputs. This path makes the paper's evidence surface reviewable without granting access to raw production exports. It also makes negative findings actionable: a count mismatch, a broken join, a leaked token, or an unreproducible verifier output points to a specific release gate rather than to an opaque preprocessing step.

The current private staging audit supports that boundary. HIKARI staging contains the source exports and challenge material from which the eight JSON families were derived; SOCpilot staging contains 200 raw incident exports with operational fields that are intentionally not release artifacts. The derived side contains the 16{,}009-record HIKARI tables and the 200 canonical SOCpilot packages. We include this distinction because operational datasets often fail at exactly this point: authors either release too little, leaving only unverifiable claims, or release too much, exposing sensitive context. Our methodology treats the boundary as a design object with checks, not as a paragraph of assurance.

The aggregate release audit makes those checks quantitative without exporting identifiers. Across HIKARI, all \HikariRecordCount{} records conform to their source-family schema, \HikariTimestampRate{} carry a parseable timestamp, all EDR records retain lineage fields, and every record retains at least one linkable entity field. However, \HikariSingletonEntityRate{} of distinct entity values and \HikariSingletonSignatureRate{} of quasi-identifier signatures are singletons. The scan also finds \HikariEmailPatternCount{} e-mail-shaped strings, \HikariUserPathPatternCount{} user-path patterns, and \HikariGlobalIpPatternCount{} occurrences of globally routable addresses. Only \HikariOrderRate{} of adjacent parseable timestamps are nondecreasing in file order, so replay must sort explicitly rather than infer chronology from row position. These are release-blocking review triggers, not confirmed leaks. They keep the HIKARI candidate in private staging until field-level adjudication and keyed rewriting. The SOCpilot blind export independently reports \SocpilotIncidentCount{}/\SocpilotIncidentCount{} valid packages, \SocpilotEventCount{} telemetry events, \SocpilotMappedTaskCount{}/\SocpilotMappedTaskCount{} mapped tasks with no ambiguous or fallback match, and \SocpilotPrivacyIssueCount{} findings under its configured forbidden-token scan. Its four rules constrain only \SocpilotPolicyCatalogCoverage{} of the five-action catalog, bounding policy-coverage claims even though the evaluated approval rules remain auditable. The audit is therefore an empirical linkage screen, not a formal re-identification bound.

\section{Validation}
\label{sec:validation}

We validate the datasets on two fronts with different consumers (human trainees and autonomous agents), because an artifact that survives both is more credibly investigation-preserving than one tuned to a single use.

\textbf{HIKARI training front.} HIKARI~\cite{hikari} is a defensive cyber-range that runs the injected logs through a standard monitoring stack: CTFd for challenge management, Kafka for event streaming, Logstash for ingestion, and Elasticsearch with Kibana for indexing and investigation. The stack is provisioned as infrastructure-as-code behind a Virtual Private Network (VPN)-controlled access boundary. Unlike offensive Capture-the-Flag, the participant acts as a SOC analyst. Evidence is released progressively via Kafka as challenges are unlocked, so the investigation unfolds over time rather than from a static dump. From the anonymized datasets, we built \textbf{37 defensive challenges}, each derived from a real detection playbook and mapped to MITRE ATT\&CK~\cite{mitre_attack} tactics and techniques. Concrete challenge types include malicious shell and reverse shell execution, backdoor account creation, Active Directory enumeration, web exploitation attempts, suspicious \texttt{wget} and \texttt{curl} downloads, ransomware-like access patterns, lateral movement, and EDR evasion. Each is anchored to the source whose telemetry carries the evidence, so a single challenge often forces the analyst to correlate data across endpoints, identities, perimeters, or the web. A challenge is solvable only if anonymization preserves temporal coherence, consistent entity pseudonyms, interpretable technical fields, and a benign background dense enough that malicious activity must be reasoned out rather than read off a label.

HIKARI exercises the datasets at four levels. At the \emph{technical} level, the anonymized logs must be ingested, indexed, and visualized through the stack. At the \emph{investigative} level, they must preserve enough semantics for a participant to locate the evidence. At the \emph{pedagogical} level, they must sustain a coherent challenge narrative. At the \emph{operational} level, the scenarios must resemble real SOC work, such as reconnaissance, exploitation, malicious execution, lateral movement, persistence, web exploitation, or suspicious authentication, and not artificial puzzles. We use successful construction and replayable injection as evidence of preserved utility, not as a cohort-performance claim. Solving rates, time-to-flag distributions, and participant error analysis are the next layers in the measurement stack.

\textbf{SOCpilot evaluation front.} The same principles extend to incident records. We anonymize 200 real SOC incidents into canonical packages and pair each with a human analyst baseline reconstructed from the case-management workflow. Incidents enter the dataset under three criteria that enable paired comparison: a closed lifecycle, structured task availability, and detection diversity. The dataset preserves cross-category and severity diversity (Table~\ref{tab:severity}). Malware (76) and suspicious-activity (73) incidents dominate, and medium- and high-severity incidents together account for 87.5\% of incidents, an operational profile in which most cases require structured triage. SOCpilot~\cite{socpilot} treats the LLM as a non-autonomous proposer. It emits an action trace over a fixed catalog, and a deterministic verifier checks the trace against typed policy rules: mandatory, ordering, and approval-gated. Combining 200 incidents, two providers, two prompt arms, and one official execution each yields \textbf{800 completed trajectories}, each compared against its paired human baseline. Two results matter for the dataset claim. First, prompt-level policy text did not guarantee compliance; the two providers moved in opposite directions when policy was added to the prompt (Table~\ref{tab:socpilotmetrics}). Second, the verifier removed \textbf{466 non-compliant approval-gated actions without reducing task coverage} relative to the human baseline: 431 were removed by rule R3 and 35 by rule R4. No out-of-catalog action appeared as the dominant failure. The failure was subtler and more operationally relevant: the model selected valid response actions, such as isolation or restoration, in contexts where approval policy forbade immediate execution. That is exactly the operationally-unsafe-but-plausible recommendation that only a policy-aware check catches.

\begin{table}[!htb]
\caption{SOCpilot policy compliance}
\label{tab:socpilotmetrics}
\centering\footnotesize
\arenazebra
\begin{tabular*}{\columnwidth}{@{\extracolsep{\fill}}lcccr@{}}
\toprule
\textbf{Arm} & \textbf{Viol.} & \textbf{95\% CI} & \textbf{Cov.} & \textbf{$\Delta$J} \\
\midrule
Claude / zero   & 0.36 & [0.30,0.43] & 0.7567 & 0.0678 \\
Claude / policy & 0.87 & [0.82,0.91] & 0.7634 & 0.1544 \\
GPT / zero      & 0.54 & [0.47,0.61] & 0.8100 & 0.1092 \\
GPT / policy    & 0.47 & [0.40,0.54] & 0.7400 & 0.0917 \\
Human baseline  & 0.00 & [0.00,0.02] & 1.0000 & 0.0000 \\
\bottomrule
\end{tabular*}
\end{table}

The metric pattern explains why the incident dataset matters. The same 200 anonymized incidents, the same action catalog, and the same verifier expose a provider-specific policy effect without changing the cases. In one arm, adding policy text increased the violation rate; in another, it reduced it. The result does not claim that one provider is better. It shows that a reusable, paired incident dataset can distinguish prompt-level policy exposure from action-level operational compliance, a distinction that free-text response quality cannot measure.

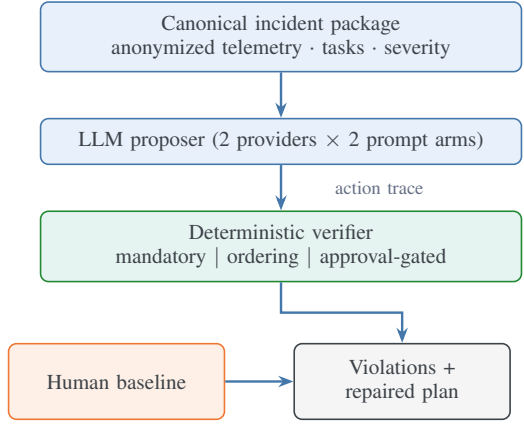
\begin{figure}[t]
\centering
\begin{tikzpicture}[arena, font=\footnotesize, node distance=6mm and 6mm]
  \node[arena node, text width=6cm] (pkg) {Canonical incident package\\\footnotesize anonymized telemetry · tasks · severity};
  \node[arena node, below=of pkg, text width=6cm] (llm) {LLM proposer \footnotesize (2 providers $\times$ 2 prompt arms)};
  \node[arena policy, below=of llm, text width=6cm] (ver) {Deterministic verifier\\\footnotesize mandatory $\mid$ ordering $\mid$ approval-gated};
  \node[arena box, draw=arenaInk, below=8mm of ver, xshift=1.6cm, text width=2.5cm, minimum height=10mm, align=center] (out) {Violations + repaired plan};
  \node[arena attack, left=9mm of out, text width=2.5cm, minimum height=10mm, align=center] (base) {Human baseline};
  \draw[arena flow strong] (pkg) -- (llm);
  \draw[arena flow strong] (llm) -- node[font=\scriptsize, text=arenaGray, fill=white,
    inner sep=1pt, pos=0.5, xshift=13mm]{action trace} (ver);
  \draw[arena flow strong] (ver.south) -- ++(0,-4mm) -| (out.north);
  \draw[arena flow strong] (base.east) -- (out.west);
\end{tikzpicture}
\caption{SOCpilot validation loop.}
\label{fig:socpilot}
\end{figure}

The artifact that crosses the privacy boundary is the canonical package, the incident-response analog of the JSON transform used for the HIKARI logs. Each package carries the minimum sufficient structure for auditable evaluation: anonymized incident metadata; detection category and severity; anonymized telemetry and evidence fields; the extracted workflow tasks; the mapped human-analyst actions; conversion and quality-control metadata; and a manifest that supports traceability. Raw exports and irreversible mappings never leave the private environment. This representation provides reproducibility (every incident follows the same protocol), comparability (the LLM plan and the human baseline share a common action vocabulary), and auditability (each violation traces to structured evidence).

\begin{table}[!htb]
\caption{SOCpilot severity profile}
\label{tab:severity}
\centering\small
\arenazebra
\begin{tabular*}{\columnwidth}{@{\extracolsep{\fill}}lrr@{}}
\toprule
\textbf{Severity} & \textbf{Incidents} & \textbf{Share} \\
\midrule
Critical    &  9 &  4.5\% \\
High        & 83 & 41.5\% \\
Medium      & 92 & 46.0\% \\
Low         & 11 &  5.5\% \\
Unspecified &  5 &  2.5\% \\
\midrule
\textbf{Total} & \textbf{200} & \textbf{100\%} \\
\bottomrule
\end{tabular*}
\end{table}

The evaluation is auditable because both the dataset and the verifier are explicit. The action catalog covers common SOC response decisions such as forensic collection, host isolation, credential reset, egress blocking, and host restoration. The human baseline is projected into this same catalog via a deterministic task-to-action mapping, so the plan and the baseline share a common vocabulary. The policy surface types each rule as \emph{mandatory}, \emph{ordering}, or \emph{approval-gated}. In addition to these, the verifier reports the run-level violation rate, task coverage relative to the paired human baseline, the enforcement-modification rate, and the Jaccard change in the action set after enforcement. Together, these determine whether a plan is \emph{compliant} rather than whether it \emph{overlaps} human practice. The metrics span the ten detection categories of the dataset (Table~\ref{tab:categories}), so the measurement reflects the breadth of investigation classes that a SOC handles, not a single alert type.

Utility preservation is checked at two levels. The end-to-end level is what HIKARI and SOCpilot exercise: can a human solve an injected challenge, and can a verifier evaluate an LLM plan against a paired baseline? The micro level asks whether specific investigative operations still work after anonymization. Entity-linking checks verify that the same host, account, or process is consistent across sources. Process-lineage checks verify that parent-child execution chains survive token replacement. Temporal checks verify that multi-stage activity remains ordered. Category checks verify that a playbook label is supported by retained fields rather than by hidden analyst knowledge. These micro-benchmarks are the natural bridge from this work to a public artifact evaluation because they turn ``utility preserved'' into operations a reviewer can rerun.

\begin{table}[!htb]
\caption{SOCpilot detection categories}
\label{tab:categories}
\centering\small
\arenazebra
\begin{tabular*}{\columnwidth}{@{\extracolsep{\fill}}lrlr@{}}
\toprule
\textbf{Category} & \textbf{N} & \textbf{Category} & \textbf{N} \\
\midrule
Malware             & 76 & Authentication     & 4 \\
Suspicious activity & 73 & Falcon detection   & 4 \\
Exploit             & 19 & Potential exploit  & 4 \\
System              &  8 & Policy             & 3 \\
Access              &  7 & Reconnaissance     & 2 \\
\bottomrule
\end{tabular*}
\end{table}

\textbf{Threats to validity.} The datasets originate from a single financial-sector SOC, which strengthens operational realism but limits direct generalization. The human baseline is a retrospective operational reference, not a universal optimum, since analysts worked under time pressure, approval flows, and the evidence available at the time. Anonymization may also remove context that the original analyst could see, and the action catalog and policy surface bound what the SOCpilot evaluation can observe. We therefore do not claim a formal re-identification bound, an independently adjudicated label-error rate, or public release of the full production-derived dataset here. These become release gates for a vetted subset, reached through frequency and linkage-risk checks, rare-event review, and a multi-analyst label audit. They scope the claim without undermining it: the datasets support auditable evaluation under a declared protocol, and they set the agenda for extending the methodology to more organizations, policies, and providers.

\section{Discussion}
\label{sec:discussion}

The two validation fronts stress different properties of the same artifact boundary. HIKARI treats the data as training material and fails \emph{loudly}: if anonymization breaks temporal order, entity consistency, or field semantics, a human investigator cannot reach the evidence. SOCpilot treats the data as a measurement substrate and fails \emph{quietly}: an LLM can emit a fluent, catalog-valid plan whose 466 approval-gated actions are still operationally non-compliant. An artifact that survives both consumers is more credibly investigation-preserving than one tuned to either alone.

This is why the release design keeps two artifacts rather than collapsing them into one. HIKARI is event-level: it tests whether anonymized telemetry still supports correlation, temporal investigation, and challenge replay. SOCpilot is incident-level: it tests whether anonymized case packages still support policy-checked response planning against a human baseline. Both share dictionaries, manifests, and private-to-public governance, but each exposes a different evidence surface and therefore a different failure mode.

\textbf{Release and utility gates.} The boundary must be measurable, not asserted. A candidate release must rebuild deterministically from a manifest, validate every source schema, pass forbidden-token and rare-token scans, and undergo linkage-risk review across sources and against public or institutional dictionaries. For network fields, the goal is not ``all IPs removed,'' because that would destroy flow semantics; synthetic addresses must preserve investigation-relevant topology while preventing recovery of the original subnet. Privacy checks alone are still insufficient: a safe artifact can be useless. Utility gates, therefore, check entity-linking precision, process-lineage completeness, temporal ordering, and dictionary-level interpretability. These gates turn ``utility preserved'' into operations a reviewer can rerun.

\textbf{Reviewer replay path.} The intended reviewer path is deliberately narrow. A reviewer should first rebuild the aggregate manifest and confirm that private staging is excluded, then verify the HIKARI family counts, schema conformance, timestamp parseability, lineage-field retention, and linkable-entity rates from the anonymized tables. Next, the reviewer runs the forbidden-token and singleton screens and should observe the same blocking result for HIKARI: high singleton rates and pattern matches require adjudication before release. Finally, the reviewer replays the SOCpilot canonical packages against the action catalog and verifier, recomputing incident count, task mapping, privacy-scan status, and policy-surface coverage. This path reproduces every quantitative claim in the paper without revealing raw exports, private dictionaries, or incident identifiers.

The negative HIKARI gate is part of the contribution, not an embarrassment to hide. A release process that always passes is not a safety mechanism; it is a formatting step. Here, the gate exposes the tension that makes production SOC telemetry both scientifically valuable and dangerous: the same entity consistency that preserves investigative utility also creates linkage risk. The correct outcome is therefore not ``release everything'' or ``redact until safe.'' It is a measured boundary: SOCpilot currently supports aggregate replay, whereas HIKARI remains a private release candidate until keyed rewriting and field-level adjudication reduce linkage risk without destroying investigative joins.

That boundary also defines the next release step. A vetted public subset should publish gate outputs and replay paths, not assurances: which fields were rewritten, which singleton patterns were adjudicated, which process-lineage and entity-linking checks still pass, and which playbook categories remain covered. Future extensions should add organizations, policies, and playbooks only after each new source can pass the same private-to-public protocol rather than by relaxing the gate.

This also defines the conditions under which the methodology fails. If a rebuilt table changes record counts, if a dictionary field no longer supports a challenge, if process lineage breaks after token replacement, if an LLM trajectory cannot be mapped to the shared action catalog, or if a policy verdict depends on hidden analyst context, the artifact has failed. Those failure modes are concrete enough for reviewers and coauthors to reproduce, which is the difference between an operational data story and a reusable scientific artifact.

\section{Related Work}
\label{sec:related}

Empirical SOC studies explain why production data is hard to reproduce: most network alerts are not true attacks~\cite{yang2024trueattacks,alahmadi2022falsepositives}, and rulesets, alerts, and incidents drift over time~\cite{vermeer2022ruling}.\footnote{The literature surveyed in this section was assembled with a reproducible, versioned venue corpus~\cite{topvenues}.} Public datasets solve adjacent problems. CICIDS, UNSW-NB15, CTU-13, LANL, and CERT support reproducible benchmarking, botnet analysis, enterprise-scale traces, or insider narratives. DARPA Transparent Computing/OpTC~\cite{darpa_tc} and Splunk BOTS~\cite{splunk_bots} move closer to incident investigation through provenance and scenario structure. However, their declared objects differ from ours. A SOC training artifact must preserve the analyst's evidence path across heterogeneous logs, and an LLM-response artifact must preserve the decision context that makes a plan safe or unsafe under policy.

Network-measurement archives provide the strongest privacy counterpoint. CAIDA removes payload and applies Crypto-PAn with documented key rotation to backbone headers~\cite{caida_traces}, while MAWI scrambles addresses in daily WIDE backbone traces~\cite{mawi_archive}. Therefore, repeatable topology-aware release is feasible; however, packet-level Internet measurement does not require the cross-source process, identity, case, and response relations that SOC investigation needs. Our problem lies between these lines: retain SOC semantics, then measure the linkage risk that retention creates, rather than assuming pseudonymization ended it.

Detection and response research guides, whose structure must survive. Semantic Advanced Persistent Threat (APT) representations and attribution knowledge graphs motivate the preservation of process, identity, temporal, and communication relations~\cite{geneapt2014,cskgapt2023}. Provenance-graph reduction and intelligent-analysis roadmaps show why those relations must remain compact enough for investigation~\cite{devilnoise2024,roadmapapt2012}. SOC studies and playbook work~\cite{kokulu2019matched,schlette2024playbooks,woods2023lessons} add that analysts operate under false positives, legal constraints, and coordination flows, so a dataset must support response sequencing, not just detection labels.

Finally, LLM-assisted response makes the artifact boundary testable. Security LLM benchmarks~\cite{deng2024pentestgpt,bhatt2024cyberseceval} evaluate capability, but a plausible action plan can still violate policy. SOCpilot~\cite{socpilot} connects enforceable-policy theory~\cite{schneider2000enforceable,ligatti2005edit} with recent privilege control for LLM agents~\cite{shi2025progent,wang2025agentspec}. Our contribution is the conversion layer that makes those evaluations possible on production-derived evidence: manifests, dictionaries, paired baselines, and deterministic verifiers let a reviewer trace source selection to anonymized record, record to task, task to baseline action, and proposed action to policy verdict. Without that chain, a realistic-looking SOC dataset remains an anecdote.

The contribution is therefore a disciplined conversion layer: production-derived telemetry is transformed into two reviewable research artifacts with declared privacy gates and downstream validation modes. That is the missing step between private SOC exports and public benchmarks: preserving operational context while making evaluation claims auditable.

\section{Conclusion}
\label{sec:conclusion}

Realistic SOC data is scarce because the realistic source, production telemetry, is hard to share. We presented a methodology that extracts, anonymizes, structures, and quality-controls production SIEM logs into reusable artifacts, and showed, through 37 HIKARI challenges and a 200-incident SOCpilot evaluation, that carefully treated production data can support training, experimentation, and LLM-response evaluation under an explicit privacy--utility boundary. The contribution is the conversion layer rather than a raw-log release or a formal anonymity proof: manifests, dictionaries, quality gates, paired baselines, and deterministic verifiers make claims traceable from source selection to anonymized records, tasks, human actions, LLM actions, and policy verdicts. Operational evidence is not automatically scientific evidence; reviewer replay, linkage-risk measurement, and utility gates are what make it reusable.

\bibliographystyle{IEEEtran}

\bibliography{references}

\end{document}